\documentclass[10pt,conference]{IEEEtran}
\usepackage{color}
\usepackage{amsmath}
\usepackage{amssymb}
\usepackage{graphicx}

\IEEEoverridecommandlockouts   

\begin{document}

\title{
Rewritable Codes for Flash Memories Based Upon Lattices, and an Example Using the E8 Lattice
}

\author{\IEEEauthorblockN{Brian M.~Kurkoski} 
\IEEEauthorblockA{University of Electro-Communications\\ 
Tokyo, Japan\\ 
kurkoski@ice.uec.ac.jp} 
\thanks{This research was supported in part by the Ministry of Education, Science, Sports and Culture; Grant-in-Aid for Scientific Research (C) number 21560388.}%
\thanks{This work has been submitted to the IEEE for possible publication. Copyright may be transferred without notice, after which this version may no longer be accessible.   This paper is available on ArXiV. Submitted to Globecom 2010.}
}


\renewcommand{\thefootnote}{\hfill}	
\renewcommand{\v}{\mathbf}

\maketitle

\begin{abstract}
A rewriting code construction for flash memories based upon lattices is described.  The values stored in flash cells correspond to lattice points.  This construction encodes information to lattice points in such a way that data can be written to the memory multiple times without decreasing the cell values.  The construction partitions the flash memory's cubic signal space into blocks.   The minimum number of writes is shown to be linear in one of the code parameters.  An example using the E8 lattice is given, with numerical results.
\end{abstract}

\section{Introduction}

Rewriting codes are a coding-theoretic approach to allow rewriting to memories which have some type of write restriction, typically values stored in memory may only be increased.  While codes for binary media were proposed in the 1980s \cite{Rivest-infctrl82}, \cite{Cohen-it86}, within the past few years, a large number of rewriting codes directed at flash memory have been described 
\cite{Jiang-isit07}, 
\cite{Bohossian-isit07}, 
\cite{Yaakobi-aller08}, 
\cite{Finucane-aller08}, 
\cite{Mahdavifar-isit09}, \cite{Jiang-pacrim09}.
Most of these these codes are designed for flash memory cells that can store one of $q$ discrete levels, where the values can only increase on successive rewrites.   

However, in the physical flash cell, charge is stored during write operations.  Charge, read as a voltage, is an inherently continuous quantity. Commercial flash memory integrated circuits use analog-to-digital conversion, and present $\log_2 q$ bits per cell of digital data externally.    Currently, any coding, for error-correction and rewriting, must operate on these discrete values.  However, one might expect that future coding schemes may have access to the continuous, or analog values stored in the flash memory cells.

This paper describes a rewriting code based upon lattices, and assumes that the analog values are available for coding.   The values stored in flash cells correspond to lattice points.    From a lattice perspective, conventional rewriting codes stores data at the points $\{0,\ldots,q-1\}^n$, in a rectangular lattice.  However, rectangular lattices are inefficient, and there exist lattices that have many desirable properties such as better packing efficiency.

Because the flash cell values are continuous quantities, this paper takes the signal-space viewpoint that has long been used for the AWGN channel.  Among other results, it is now known that lattices can achieve the capacity of the AWGN channel \cite{Loeliger-it97} \cite{Erez-it04}, and lattices appear to be a promising practical approach for bandwidth-constrained channels \cite{Sommer-it08}. In fact, a related technique, trellis-coded modulation, has already been considered for error-correction in flash memories \cite{Sun-CSD07}.    In this paper, error-correction is not explicitly considered, however it is an important aspect of using lattices in flash memories.   

An important consideration for both flash memories and AWGN channels is the power constraint.  For AWGN channels, the average power constraint induces an ideally spherically-shaped codebook, which can be well-approximated by a shaping region equal to the Voronoi region of high dimensional lattices.   However, encoding in a shaping region requires computationally expensive lattice quantization \cite{Erez-it04}.   But for flash memories, the ``peak" power constraint is cubic, that is, all points are within the cube $(0,q-1)^n$, corresponding to the fact that the voltage on each cell has a minimum and maximum possible value.  Fortunately, lattice quantization is not required.   As was shown by Sommer, et~al., when the lattice has a triangular generator matrix, there is an efficient encoding which results in a cubic shaping region \cite{Sommer-itw09}.   This paper presents a slight generalization of this method.

The proposed code partitions the signal space into $D^n$ blocks with maximum volume $M^n$; these terms will be defined in the next section, but one may assume $D\cdot M = q-1$.  Some, but not necessarily all blocks, have a one-to-one mapping between information and lattice points contained within that block.  When the memory is to be rewritten with new information, a new codeword either within the same block, or in an adjoining block, is selected, such that the cell values are only increased.   While there are multiple codeword candidates that can encode the new information, the codeword which maximizes the future number of rewrites is selected.

\begin{figure*}
\begin{center}
\includegraphics[width=16cm]{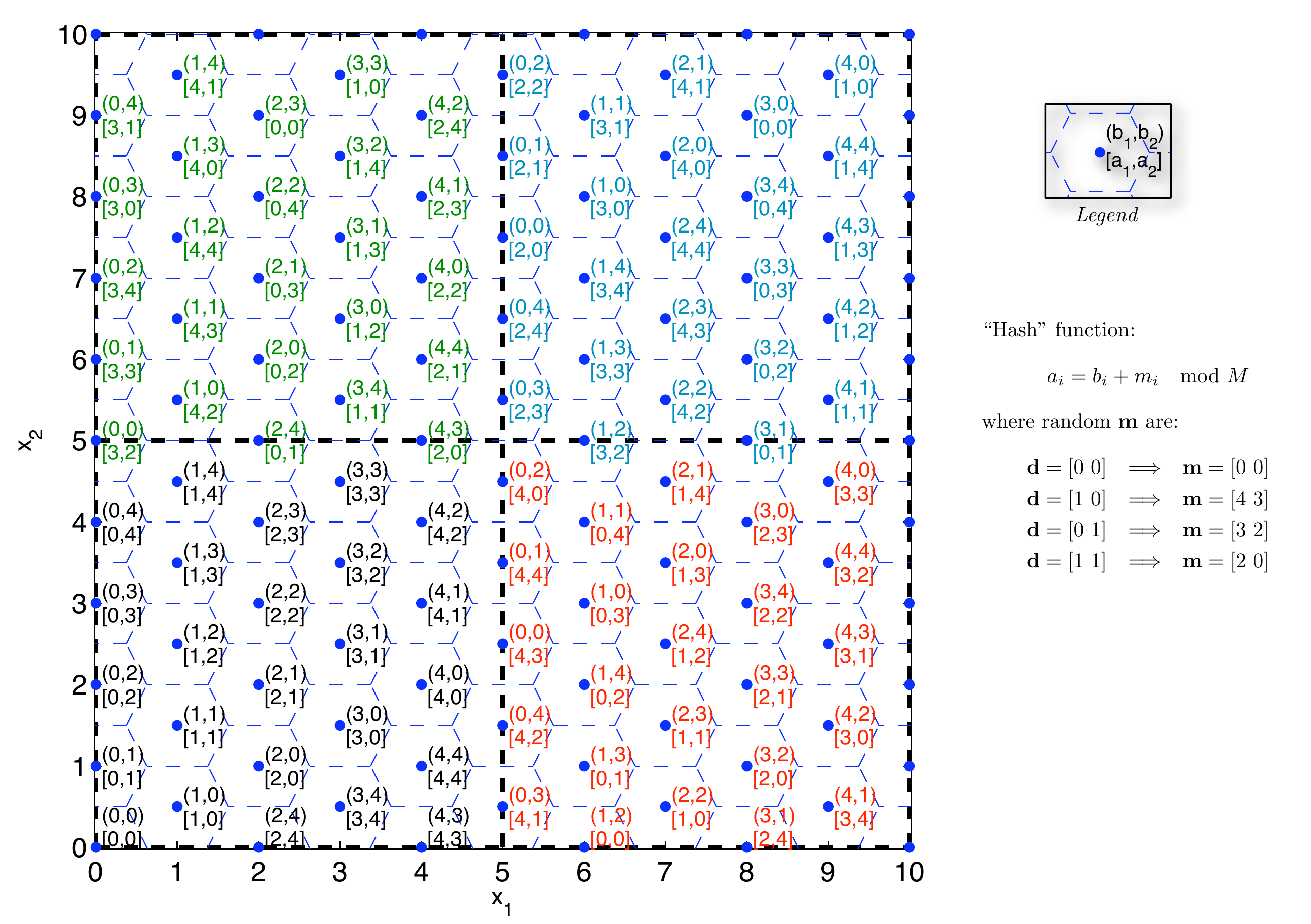}
\end{center}
\caption{Illustration of the proposed code for two dimensions, $n=2, G= [ 1\  0 ; \frac 1 2\  1], M=5, D=2$}
\label{fig:fig1}
\end{figure*}

The triangular-generator lattice encoding is linear, but unfortunately the linearity presents a problem.   When new data is to be written to the memory, under a linear construction there is exactly one new codepoint nearest the current state codepoint.   But if at least one component of the current state is near the boundary, then with some probability, the nearest ``codepoint" will be a phantom, outside of the power constraint.   It might be acceptable to select another, suboptimal codepoint.  But because of linearity, all such codepoints are phantoms and unaccessible.   Accordingly, a random ``hash" is introduced.   This destroys the linearity, and requires a procedure to select the candidate codeword which is most suitable for rewriting.   But its purpose is to increase the average number of times the memory may be written.

While rewriting codes for flash memories have received some research attention, error-correction coding for flash memories is of considerable practical importance \cite{Micheloni-ISSCC06} \cite{Chen-08}.   There have been only a few studies on the dual-purpose codes which can both correct errors and allow rewriting \cite{Zemore-it91} \cite{Jiang-isit07*2}.   However, the simple concatenation of a rewriting code and error-correction code appears to be problematic. Encoding the rewriting code followed by a systematic error-correction code means that parity bits are not rewritable.  On the other hand, switching the concatenation results in no guarantees of minimum distance, since most rewriting codes do not appear to be systematic.  However, lattices considered in this paper have a natural error-correction property, due do the Euclidean distance that separates the points.  While this paper assumes there is no noise, the goal is to show that a rewriting code can be constructed by an appropriate encoding from information to lattice points.   

\section{Code Construction}

\subsection{Lattices}

An $n$-dimensional lattice $\Lambda$ is defined by an $n$-by-$n$ generator matrix $G$.   The lattice consists of the discrete set of points $\mathbf x = (x_1,x_2, \ldots, x_n)^t$ for which 
\begin{eqnarray}
\mathbf x &=& G \mathbf b, \label{eqn:lattice}
\end{eqnarray}
where $\mathbf b =(b_1,\ldots, b_n)^t$ is from the set of all possible integer vectors, $b_i \in \mathbb Z$.  The Voronoi region is region of $\mathbb R^n$ which is closer to $\v x$ than to any other point, and the volume of this region is the determinant of $G$:
\begin{eqnarray}
V(\Lambda) &=& | \det G | .
\end{eqnarray}
The $i,j$ entry of $G$ is denoted $g_{ij}$.

\subsection{Codebook}

Let $\Lambda$ be a lattice with a diagonal generator matrix.   Let $B$ be an $n$-cube, given by:
\begin{eqnarray}
& 0 \leq x_i < D \cdot M,
\end{eqnarray}
for $i=1,\ldots,n$, which has volume $(D M)^n$.   Then the codebook of the proposed code is:
\begin{eqnarray}
\mathcal C = \Lambda \cap B.
\end{eqnarray}
The lattice generator is lower triangular, and the diagonal entries $g_{ii}$ satisfy the condition that $M / g_{ii}$ is an integer.

The cube $B$ is partitioned into $D^n$ blocks.  The blocks are indexed by $\mathbf d$, given by:
\begin{eqnarray}
\mathbf d &=& \{d_1, d_2, \ldots, d_n\}, \textrm{  with } d_i \in \{0,1,\ldots, D-1\}.
\end{eqnarray}
Each block $B_{\mathbf d}$ is given by the set of $\mathbf x \in \mathbb R^n$ such that:
\begin{eqnarray}
d_i M \leq x_i  < (d_i + 1) M \textrm{ and } x_i < D M,
\end{eqnarray}
for $i=1,\ldots, n$.  If $D$ is an integer, then then each block is an $n$-cube with volume $M^n$.  However, $D$ is allowed to be non-integer, in which case some blocks sharing a face with $B$ will have volume less than $D^n$.

The lattice points inside each block form a subcodebook:
\begin{eqnarray}
\mathcal C_{\mathbf d} &=& \Lambda \cap B_{\mathbf d}.
\end{eqnarray}
The size of the full codebook and the maximum size of any subcodebook are
\begin{eqnarray}
\frac{
(D\cdot M)^n
}{
| \det G |
}
& \textrm{and}
&
\frac{
M ^n
}{
| \det G |
},
\end{eqnarray}
respectively.

Within each block with volume $D^n$, there is a one-to-one mapping from information to subcodewords, thus the rate of the code, expressed in information bits per cell is:
\begin{eqnarray}
R &=& \frac{
\log_2  \big( M ^n /  | \det G | \big)
}{
n
}
= \log_2 M,
\end{eqnarray}
if $| \det G | = 1$ is used. Also, there is a one-to-many, in particular, a one-to-$D^n$ mapping between information and the full codebook.   

\subsection{Encoding}

The encoding is as follows, and is illustrated in Fig.~\ref{fig:fig1} for $n=2$.  A random ``hash'' maps information $\mathbf u = (u_1,\ldots, u_n)$ to hashed sequences $\mathbf a = (a_1, \ldots, a_n)$.   This hash depends upon $\mathbf d$:
\begin{eqnarray}
h(\mathbf d) : \v u \rightarrow \v a
\end{eqnarray}
A simple hash is simply to add a constant modulo $M$:
\begin{eqnarray}
a_i &=& u_i + m_{i,\v d} \mod M,
\end{eqnarray}
where $m_{i,\v d}$ is a hash vector for block $\v d$.

These symbols are then encoded to lattice points as
\begin{eqnarray}
\mathcal E : \mathbf a \rightarrow \mathbf x,
\end{eqnarray}
where $\v x \in \mathcal C$.

The encoding $\mathcal E$ for any block $\v d$ is as follows \cite{Sommer-itw09}.   In general, $G \cdot \mathbf a$ is not in $B_{\mathbf d}$.  Instead, the encoding finds
\begin{eqnarray}
\mathbf b &=& \mathbf a + M \mathbf k,
\end{eqnarray}
with $\v k = (\frac{k_1}{g_{11}}, \ldots, \frac{k_n}{g_{nn}})$ such that 
\begin{eqnarray}
\mathbf x = G \cdot \mathbf b
\end{eqnarray}
is in the cube $B_{\mathbf d}$.  Because the generator matrix is diagonal, the $k_i$ can be found by solving the inequality:
\begin{eqnarray}
 d_i M \leq  & x_i  & < (d_i+1) M \\
d_i M \leq & \sum_{j=0}^{i-1} g_{ji} b_j + g_{ii} \big( a_i + \frac{M}{g_{ii}} k_i ) & <  (d_i+1) M
\end{eqnarray}
for $k_i$, which is unique.   First $k_1$, then $k_2, \ldots k_n$ are found in sequence.   In particular:
\begin{eqnarray}
k_i &=& \Big\lceil \frac{
d_i M - \sum_{j=0}^{i-1} g_{ji} b_j - g_{ii} a_i
}{
M 
}
\Big\rceil,
\end{eqnarray}
where computation at step $i$ depends upon $b_1,\ldots, b_{i-1}$.
Also, the data range depends on $g_{ii}$, that is $a_i \in \{0,1,\ldots, \frac M {g_{ii}}-1\}$.

\renewcommand{\c}[1]{| #1 |^+}

Now, consider that the current state of the memory is $\v s$.   Given an information sequence $\v u$, or its hash $\v a$, there may be many candidate codewords.  For any codeword $\v x$, all components of $\v x - \v s$ must be positive. Let $\mathbf x [\mathbf d]$ denote the codeword in $\mathcal C_{\mathbf d}$ corresponding to $\v u$.

Since there is no a priori knowledge about future data sequences, it is reasonable that the codeword choice should maximize the number of codeword points that remain ``available" to future writes, that is, the number of codewords in the positive direction should be maximized.   While it is computationally difficult to count these points, a reasonable approximation is the volume that remains after the point is written.  This argument bears some resemblance to the continuous approximation used in channel coding using lattices \cite{Forney-it88}.  In particular, if $\v x$ is to be written, then the remaining volume is:
\begin{eqnarray}
\prod_{i=1}^n \big( M\cdot D - x_i\big)
\end{eqnarray}
and the encoder should write:
\begin{eqnarray}
\v x &=& \max_{\v d :( \v s - \v x[\v d]) \geq \v 0} \prod_{i=1}^n \big( M\cdot D - x_i\big) .
\end{eqnarray}
This maximization is computationally complex as the lattice dimension $n$ increases.  Generally, however there will be a codeword in a neighboring block.  Thus, the search can be performed not over all $\v d$, but only over those positive neighbors of the block that contains the current state $\v s$.   This results in complexity proportional to $2^n$.

\subsection{Decoding}

\newcommand{\wv}[1]{\widehat{\mathbf {#1}}}

Decoding is straightforward.   If noise is present, then lattice decoding should be performed, to obtain the estimated lattice point $\wv{x}$.  

The encoded integers are simply $\wv{b} = G^{-1} \wv{x}$, and from these, $\wv a$ is obtained as:
\begin{eqnarray}
a_i &=& b_i \mod{M}, \textrm{for } i = 1,\ldots,n.
\end{eqnarray}
The information is obtained by reversing the hash function:
\begin{eqnarray}
u_i &=& a_i - m_{i,\v d} \mod{ \frac{M}{g_{ii}}},
\end{eqnarray}
where $m_{i,\v d}$ is defined as before.

\begin{figure}
\begin{center}
\includegraphics[width=9.5cm]{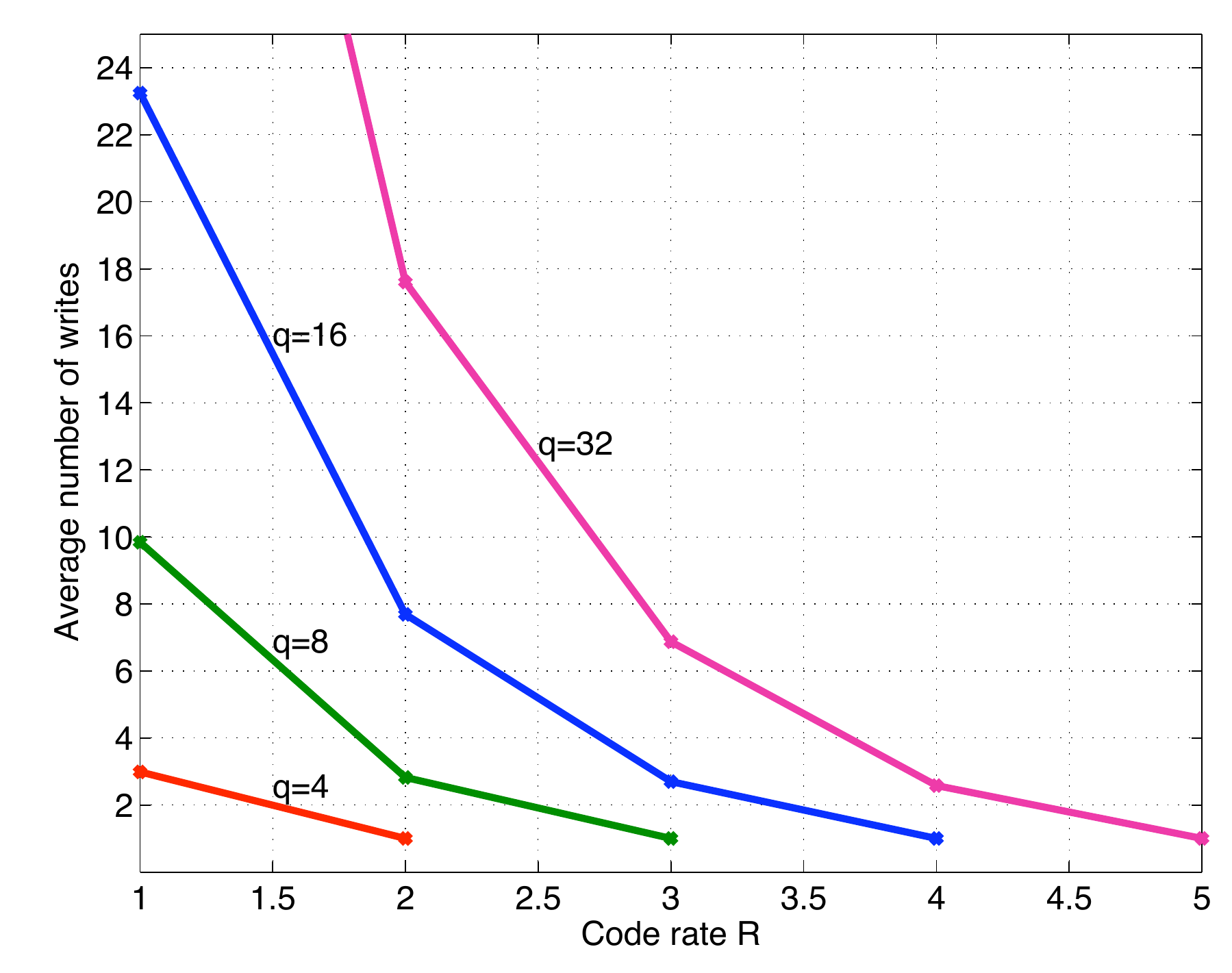}
\end{center}
\caption{Average number of {word} writes using the E8 lattice, with $q-1 = D M$ and code rate $R=\log_2 M$. }
\label{fig:fig2}
\end{figure}

\section{Numerical Results}

In order to make a fair normalization in the absence of noise, the scale of the lattice must be selected.  Proper scaling of the lattice for comparison of coding gain is not clear, although previous work on channel coding used a fixed volume of the Voronoi region (see, for example, \cite{Tarokh-it99*2}).  For conventional $q$-ary rewriting codes, the rectangular lattice with integer spacing applies; the volume of the Voronoi region of this lattice is 1.  That is, a scalar $\alpha$ is selected such that:
\begin{eqnarray}
| \det \alpha G | = \alpha^n | \det G | = 1.
\end{eqnarray}

It should be fairly clear that the minimum number of guaranteed writes is $D$.  In the worst-case scenario, a codeword is written in block $\v d = [0 \ldots 0]$ followed by $\v d = [1 \ldots 1]$ until $\v d = [D-1 \cdots D-1]$.  This may be visualized in Fig.~\ref{fig:fig1} by first writing a codeword near the upper-right-hand corner of block $[0\  0]$, and then $[1\ 1]$.  

\def\slantfrac#1#2{\tiny {\hbox{$\,^#1\!/_#2$}}}

To evaluate the average number of writes, the E8 lattice is used.  This lattice, with dimension $n=8$, has good packing properties, as well as an efficient decoding algorithm \cite{Conway-it83}, and one possible generator is:
\begin{eqnarray}
G &=&
\left[
\begin{array}{rrrrrrrr}
\slantfrac{1}{2} & 0 & 0 & 0 & 0 & 0 & 0 & 0 \\
\slantfrac{1}{2} & 1 & 0 & 0 & 0 & 0 & 0 & 0 \\
\slantfrac{1}{2} & -1 & 1 & 0 & 0 & 0 & 0 & 0 \\
\slantfrac{1}{2} & 0 & -1 & 1 & 0 & 0 & 0 & 0 \\
\slantfrac{1}{2} & 0 & 0 & -1 & 1 & 0 & 0 & 0 \\
\slantfrac{1}{2} & 0 & 0 & 0 & -1 & 1 & 0 & 0 \\
\slantfrac{1}{2} & 0 & 0 & 0 & 0 & -1 & 1 & 0 \\
\slantfrac{1}{2} & 0 & 0 & 0 & 0 & 0 & -1 & \phantom{-}2 \\
\end{array}
\right]
\end{eqnarray}
It has a lower-triangular form, and so it is suitable for the proposed construction.  
  
Naturally, there is a tradeoff between code rate and the average number of writes, and this is demonstrated in Fig.~\ref{fig:fig2}, obtained by computer simulation.   Values of $q$ were fixed, with $q-1 = DM$.   The code rate $R = \log_2 M$, and $D$ was allowed to be a non-integer.   The most striking feature is that the number of writes depends strongly upon $q$.  Also, while not shown here, it was observed numerically that the average number of writes increased roughly linearly in $D$, much as the minimum number of writes is also linear in $D$.

Note that many conventional $q$-ary rewriting codes allow rewriting one \emph{bit} at a time.   For this lattice-based code, the entire \emph{word} is re-written.

\section{Discussion}

This paper has demonstrated that rewriting codes based upon lattices is feasible.  State-of-the-art has flash chips provide digital data to the external interface, but for lattices to be applicable, the analog values should be accessible.  One of the goals of this work is to show the benefits of integrating the analog signal processing and coding in flash memories.

Lattices have an inherent error-correction property, and they appear to be suitable for both error correction and rewriting.  In fact, the equal-Voronoi-volume assumption substantially favors lattices with regard to error correction, since it is well known that increasing the dimension leads to substantial error-correction coding gain.  A point to note is that the rewriting capability of lattices presented in this paper does not appear to substantially depend upon the dimension $n$.  That is, the minimum number of writes is $D$, and there is a well-defined relationship between $R$, $D$, $q$ and $M$.   However, this appears to not be surprising.   In 1984, Fiat and Shamir, working with very general memory models, those based upon directed acyclic graphs (DAG), observed:
\begin{quotation}
\noindent The significant improvement in memory capability is linear with the DAG depth.  For a fixed number of states a ``deep and narrow'' DAG cell is always preferable to a ``shallow and wide'' DAG cell. \cite{Fiat-it84}
\end{quotation}
That is, a deep cell has a large value of $q$, and a narrow cell has small $n$.

The lattice-based construction does have one weakness when the dimension is small.  While the conventional $q$-ary construction can write the maximum value of $q-1$ in all cells, this is not possible using lattices, and leads to a slight loss of capability. Note in Fig.~\ref{fig:fig1} that there are some lattice points on the boundary which cannot be assigned to a subcodebook. However, this loss can be readily recovered by the superior packing density of lattices, obtained as the dimension increases.

Low-density lattices codes are high-dimensional lattices which can approach the asymptotic bounds for \emph{coding} gain \cite{Sommer-it08}.   These lattices are highly suitable for coding for flash, because some such lattices have a triangular generator matrix \cite{Sommer-itw09}, suitable for rectangular shaping.   Their belief-propagation decoding algorithm appears suitable for decoding in the presence of noise, including some reduced-complexity decoding algorithms \cite{Kurkoski-isit09}.


\bibliographystyle{ieeetr}

\end{document}